\documentclass[twocolumn,amsmath,amssymb,superscriptaddress,longbibliography,nofootinbib]{revtex4-1}
\pdfoutput=1

\usepackage[T1]{fontenc}
\usepackage{physics}
\usepackage{siunitx}
\usepackage{graphicx}
\usepackage[citecolor=blue, urlcolor=blue, linkcolor=blue, colorlinks=true]{hyperref}

\newcommand{\Id}{\mathbb{I}}

\begin{document}
\title{Finite-depth scaling of infinite quantum circuits for quantum critical points}
\author{Bernhard Jobst}
\email{bernhard.jobst@tum.de}
\affiliation{Department of Physics, TFK, Technische Universit{\"a}t M{\"u}nchen, James-Franck-Stra{\ss}e 1, D-85748 Garching, Germany}
\author{Adam Smith}
\affiliation{School of Physics and Astronomy, University of Nottingham, Nottingham, NG7 2RD, United Kingdom}
\affiliation{Centre for the Mathematics and Theoretical Physics of Quantum Non-Equilibrium Systems, University of Nottingham, Nottingham, NG7 2RD, United Kingdom}
\author{Frank Pollmann}
\affiliation{Department of Physics, TFK, Technische Universit{\"a}t M{\"u}nchen, James-Franck-Stra{\ss}e 1, D-85748 Garching, Germany}
\affiliation{Munich Center for Quantum Science and Technology (MCQST), D-80799 Munich, Germany}
\date{\today}

\begin{abstract}
    The scaling of the entanglement entropy at a quantum critical point allows us to extract universal properties of the state, e.g., the central charge of a conformal field theory. With the rapid improvement of noisy intermediate-scale quantum (NISQ) devices, these quantum computers present themselves as a powerful tool to study critical many-body systems. We use finite-depth quantum circuits suitable for NISQ devices as a variational ansatz to represent ground states of critical, infinite systems. We find universal \emph{finite-depth scaling} relations for these circuits and verify them numerically at two different critical points, i.e., the critical Ising model with an additional symmetry-preserving term and the critical XXZ model.
\end{abstract}

\maketitle

\section{Introduction}
The collective behavior of strongly correlated many-body systems can be very distinct from the behavior of their elementary constituents, giving rise to emergent phases of matter and transitions between them~\cite{Anderson1972}. At zero temperature, a many-body system can undergo a continuous quantum phase transition as a parameter in the Hamiltonian is varied. At the critical point, the system exhibits universal behavior, independently of the microscopic details of the system~\cite{Sachdev2011}.

In one-dimensional quantum systems, critical points are often conformally invariant, and can be described by a $1+1$-dimensional conformal field theory (CFT)~\cite{DiFrancesco1997, Calabrese2006}. CFTs can be used to obtain many analytical results for these critical systems~\cite{DiFrancesco1997, Calabrese2006, Holzhey1994, Calabrese2004, Calabrese2009}. One of the most significant findings of CFTs is that these critical points can be classified in terms of their central charge $c$. For example, for the Ising universality class, the associated central charge is $c=\frac{1}{2}$. The central charge is directly related to physical quantities such as the entanglement entropy. Consider a bipartition of the system into two halves $A$ and $B$, with reduced density matrices $\rho_A$ and $\rho_B$. The von Neumann entropy of subsystem $A$ is then defined as ${S_A = -\Tr\rho_A\log\rho_A}$ and analogously for subsystem $B$. Close to a conformal critical point, where the correlation length $\xi$ is large, the entropy scales as ${S \sim \frac{c}{6}\log\xi}$~\cite{Holzhey1994, Calabrese2004}.

To study the universal critical behavior of a given microscopic model, numerical simulations are important~\cite{Sandvik2010}. A straightforward approach is to consider the Hamiltonian of a finite system and diagonalize it exactly with a computer. However, only small system sizes are tractable with this approach because the dimension of the Hilbert space grows exponentially with the size of the system~\cite{Sandvik2010}. As the characteristic signatures of a phase transition can, strictly speaking, only occur in infinite systems, the limitation to small systems obscures the observation of these signatures in computer simulations.

\begin{figure}[t]
    \centering
    \includegraphics[width=0.966\linewidth]{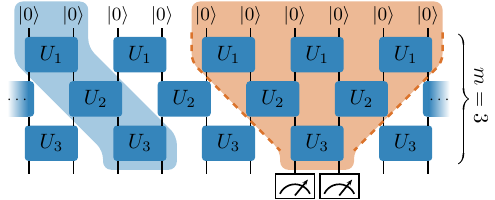}
    \caption{\textbf{Infinite brick wall quantum circuit.} The initial state is at the top, then $m=3$ layers of two-qubit gates in a brick wall pattern are applied. The infinitely repeating unit cell is shaded in light blue. The structure of the circuit leads to a causal light cone, indicated by the orange dashed line bounding the orange area. When measuring the two qubits indicated by the measurement symbol, only the gates in the orange area, i.e., the gates within the light cone, contribute, and the other gates can be ignored.}
    \label{fig:circuit}
\end{figure}

A successful numerical method in one dimension, that overcomes the problem of the exponentially growing Hilbert space, is the use of matrix-product states (MPSs)~\cite{Fannes1992, Cirac2021} as a variational ansatz and their multitude of complementary algorithms~\cite{White1992, White1993, Schollwock2005, Schollwock2011, Vidal2007, Haegeman2011, Haegeman2016, tenpy}. MPSs make simulations of large systems possible because they efficiently approximate weakly entangled ground states~\cite{Cirac2021, Schollwock2011}. Furthermore, they allow us to directly work with infinite states. MPSs follow an area law for the entanglement entropy---that is, if the system is cut into two subsystems, then the entanglement entropy scales only with the size of the boundary of the subsystem and not with the size of the subsystem itself---so the MPS ansatz is limited by the amount of entanglement it can support. For representing ground states of gapped, local Hamiltonians, this is ideal because these ground states follow the same area law~\cite{Hastings2007, Schuch2008, Eisert2010, Gottesman2010, Cirac2021}. Ground states of critical systems, however, violate the area law; the entanglement entropy diverges logarithmically at a critical point~\cite{Holzhey1994, Calabrese2004}. Therefore, MPSs cannot exactly represent critical states. Trying to represent critical states with MPSs leads to systematic deviations that are described quantitatively by the theory of \emph{finite-entanglement scaling}~\cite{Tagliacozzo2008, Pollmann2009}; using this framework, information about the critical state can still be extracted, and MPSs can be used for studying critical points~\cite{Tagliacozzo2008, Pollmann2009, Pirvu2012}.

A promising tool for studying quantum many-body systems is the use of quantum computers. As the current noisy intermediate-scale quantum (NISQ) devices improve in quality, they might very soon yield a potential speed-up compared with classical computers~\cite{Boixo2018, Arute2019}. On a quantum computer, we can use variational quantum eigensolvers (VQEs) with finite-depth circuits to represent the ground state of a given Hamiltonian or to simulate the time evolution~\cite{Peruzzo2014, McClean2016, Sweke2020, BravoPrieto2020, Lin2021, Barratt2021}. When using these finite-depth circuits for approximating the ground states of critical systems, the question arises whether similar scaling relations to the finite-entanglement scaling of MPSs exist, which allow us to extract results for the critical state from the finite-depth circuit.

To make the relation to MPSs more concrete, note that finite-depth circuits can be seen as a subset of MPSs~\cite{Gopalakrishnan2019, BravoPrieto2020, Barratt2021}. However, the circuits used in this paper (see Fig.~\ref{fig:circuit}) need exponentially fewer parameters to represent a state than a generic MPS with a similar entanglement entropy. In other words, these circuits can generate more entanglement than MPSs with the same number of parameters. This had previously been discussed for a similar circuit structure in the context of time evolved states~\cite{Schoen2005, Lin2021, Barratt2021}. In the context of ground states, this was much less discussed in the literature before. However, as we show in this paper, the circuits we use provide a systematic scaling approach to critical states and can thus be used to study critical systems. The reduction in parameters is moreover crucial for the use of NISQ devices, as the depth of circuits that can be successfully run is restricted.

In this paper, we find scaling relations for finite-depth circuits of infinite systems at quantum critical points. In Ref.~\cite{BravoPrieto2020}, authors studied the scaling of the entanglement entropy of such circuits at critical points in finite systems. The authors distinguished two regimes: the \emph{finite-depth} regime for circuits that are shallow compared with the system size, where the scaling of the entanglement entropy depends strongly on the depth of the circuit, but not on the size of the system; and the \emph{finite-size} regime for circuits that are deep compared with the system size, where the scaling depends strongly on the system size, but not on the circuit depth. For infinite systems, we observe scaling relations for the entanglement entropy consistent with those found in the finite-depth regime. Moreover, we consider other universal quantities at the critical point and can describe their scaling quantitatively. For the transverse-field Ising model in particular, authors in Ref.~\cite{Dreyer2021} could derive some scaling relations for finite-depth circuits analytically.

The paper is structured as follows: In Sec.~\ref{sec:brick_wall_circuits}, we introduce brick wall circuits as a variational ansatz and study the Ising phase transition as an example. Then in Sec.~\ref{sec:finite-depth_scaling}, we adapt the finite-entanglement scaling of MPSs to the finite-depth scaling of the brick wall circuit. Afterward, we show evidence for the finite-depth scaling in two numerical examples in Sec.~\ref{sec:numerical_evidence}. We consider the transverse-field Ising model with an additional symmetry-preserving term and the XXZ model. Finally, in Sec.~\ref{sec:discussion}, we discuss the results of the paper.

\section{Brick Wall Circuits}
\label{sec:brick_wall_circuits}
Figure~\ref{fig:circuit} shows a brick wall circuit with three layers. The initial state of the quantum computer, with all qubits in the state $\ket{0}$, is located at the top. Then three layers of two-qubit gates are applied alternatingly on even and odd bonds, creating a circuit structure that resembles a brick wall. When using the circuit as a variational ansatz, the number of layers $m$ in the circuit can be varied to change the number of parameters in the circuit. This type of circuit has been used as a variational ansatz for finite systems in Refs.~\cite{BravoPrieto2020, Haghshenas2021}. Brick wall circuits for infinite systems have been considered in the context of numerical time-evolution of the entanglement spectrum in Ref.~\cite{Gopalakrishnan2019}. A big advantage of this circuit is that its depth is independent of the number of qubits in the circuit. This makes the circuit perfect for NISQ devices, where the noise limits the number of gates we can apply successively.

The structure of the circuit in Fig.~\ref{fig:circuit} yields a causal light cone. If we perform a measurement on the two qubits marked by a measurement symbol in the figure, then only the gates within the orange shaded area contribute to the outcome. This indicates the light cone of the circuit. As only a subset of the gates contributes to a measurement outcome, it is possible to describe infinite states with a brick wall circuit. To do this, consider the two-site unit cell of the circuit, shaded in light blue in Fig.~\ref{fig:circuit}; by repeating it indefinitely, we obtain an infinite state that is invariant under translations by an even number of sites. Note that the depth of the circuit still remains the same for the infinite state. Whenever we are interested in a measurement outcome of the state, it is enough to consider the gates within the light cone to get results for the infinite state. The light cone has another consequence in that it limits the range of the correlations in the circuit. If we measure qubits that are sufficiently far apart, such that their light cones do not overlap, then their measurement outcomes are independent of one another. Consider a two-point correlation function $C(i,j) = \expval{O^A_iO^B_j} - \expval{O^A_i}\expval{O^B_j}$, where the two operators $O^A_i$ and $O^B_j$ act on sites $i$ and $j$, respectively. Then, if the brick wall circuit has $m$ layers, the correlation function $C(i, j) = 0$ if $\left|j-i\right|\geq 2m$. This means the correlation length of the circuit must be bound by $\xi < 2m$.\footnote{If the operator $O^A_i$ acts on the second site of a unit cell, the operator $O^B_j$ acts on the first site of a unit cell and $i<j$, then the correlation function $C(i,j)$ already vanishes if $j-i=2m-1$. Nevertheless, the correlation length of the circuit can scale at most linearly with $m$.}

In the following, we will treat the brick wall circuit as a variational ansatz by parametrizing the gates and optimizing them such that the circuit minimizes the expectation value of a given Hamiltonian. In principle, we could calculate the gradient of the energy with respect to the circuit on a quantum computer and perform quantum-classical optimization like VQEs~\cite{Peruzzo2014, McClean2016, Sweke2020}. However, due to the noise in current quantum computers, we use automatic differentiation to calculate the gradient on a classical computer~\cite{tensorflow, Baydin2018}. We then update the gates of the circuit using gradient-based optimization algorithms that make use of the Riemannian geometry of the manifold of unitary matrices to keep the gates unitary~\cite{qgopt, Luchnikov2021, Absil2008, Becigneul2019, Li2020}. Using this approach, we limit ourselves to circuits with maximally eight layers because optimizing the parameters takes increasingly long for deeper circuits.

\subsection*{The transverse-field Ising model}
As a concrete example, we consider the transverse-field Ising model with the Hamiltonian:
\begin{equation}
    H = \sum_i -X_iX_{i+1} + g Z_i.
\end{equation}
As we tune the parameter $g$, which controls the strength of the transverse field, we can observe a phase transition at $g=1$. For $g<1$, the system is in a symmetry-broken state with a nonzero magnetization $\expval{X_i}$, which is acting as the order parameter; for $g>1$, the ground state is symmetric, and the magnetization is zero.

\begin{figure}
    \centering
    \includegraphics[width=\linewidth]{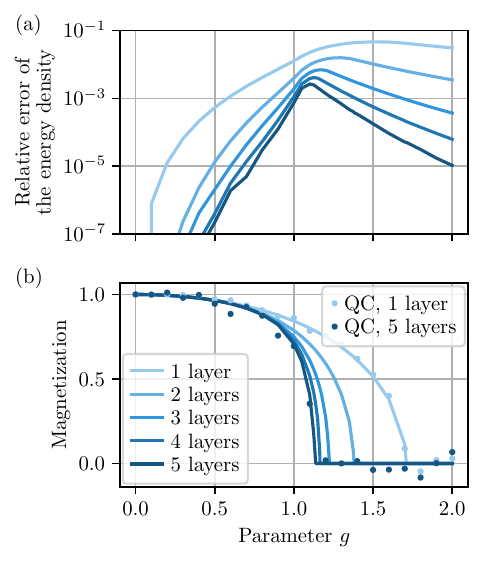}
    \caption{\textbf{Phase transition of the transverse-field Ising model as approximated by brick wall circuits.} (a) The relative error of the energy density $\left|E_{\text{opt}} - E_0\right|/\left|E_0\right|$ that the brick wall circuits achieve as a variational ansatz. Here, $E_0$ is the known ground state energy density of the transverse-field Ising model, and $E_{\text{opt}}$ is the energy density of the circuit after the optimization. (b) The local magnetization $\expval{M} = \frac{1}{2}\expval{X_i + X_{i+1}}$ of the circuits. The solid blue lines show the values obtained from numerically simulating the circuit; the blue dots show the results from the IBM quantum computer \texttt{montreal}~\cite{IBMQ}; for more details, see Appendix~\ref{app:error_correction}. Statistical error bars have been omitted, as they are smaller than the size of the dots.}
    \label{fig:Ising_transition1}
\end{figure}

Figure~\ref{fig:Ising_transition1} shows the numerical results of using brick wall circuits to represent the ground state of the Ising model as the field strength $g$ is tuned through the critical point. The relative error of the energy density is shown in Fig.~\ref{fig:Ising_transition1}a. The solid blue lines show the relative error for different numbers of layers. Increasing the number of layers also increases the accuracy. For all numbers of layers, the error is largest around the critical point, which is to be expected as the circuit cannot represent the diverging entanglement entropy and long-range correlations. Moving away from the critical point, the error decreases as we move closer to the product states at $g=0$ and $\infty$. There, the state can be exactly represented by a circuit with just single-qubit gates.

The data in Fig.~\ref{fig:Ising_transition1}b show the magnetization of the circuit throughout the transition. The solid blue lines show the magnetization as obtained from exact numerical simulation of the circuit, and the blue dots show the data obtained from the IBM quantum computers~\cite{IBMQ}. All data from the quantum computers that we present in this paper are corrected for readout errors~\cite{qiskit} and global depolarizing errors~\cite{Vovrosh2021}; for details, see Appendix~\ref{app:error_correction}. A sharp transition between the regime with zero magnetization and the regime with nonzero magnetization can be seen, both in the numerical data and in the data from the quantum computer. However, the transition point shifts with the number of layers toward the actual transition point at $g=1$. This effect is the finite-depth scaling due to the limited correlation length of the circuit, and we will discuss it more in depth in the next section. Note that being able to see a nonzero magnetization implies that we find a truly symmetry-broken state for any number of layers---this is in contrast to simulations on finite-size systems, where a symmetric ground state is found instead.

\begin{figure}
    \centering
    \includegraphics[width=\linewidth]{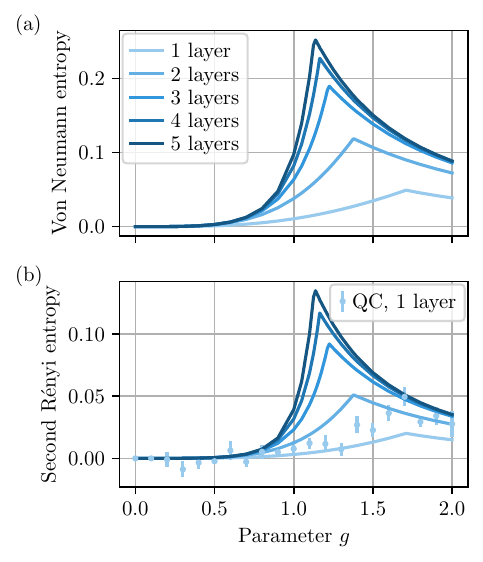}
    \caption{\textbf{Entanglement entropies of the brick wall circuits throughout the phase transition of the transverse-field Ising model.} (a) The von Neumann entropy of the brick wall circuit when used as a variational ansatz and (b) the second R\'{e}nyi entropy. The solid blue lines show the values obtained from numerically simulating the circuit; the blue dots show the average results of runs on the IBM quantum computers \texttt{montreal} and \texttt{hanoi}~\cite{IBMQ}; for more details, see Appendix~\ref{app:error_correction}.}
    \label{fig:Ising_transition2}
\end{figure}

Another way to observe the phase transition in the Ising model is to look at the entanglement entropy. If a system is cut into two parts $A$ and $B$, then the $n$th order R\'{e}nyi entropy of the subsystems is defined in terms of their reduced density matrices $\rho_A$ and $\rho_B$ as ${S^{(n)}_A = \frac{1}{1-n}\log\Tr\rho_A^n}$. In the limit $n\rightarrow1$, this reduces to the usual von Neumann entropy ${S_A = -\Tr\rho_A\log\rho_A}$. Note that $S^{(n)}_A = S^{(n)}_B$ if the full system is in a pure state, which is always the case for the circuits we consider here. The von Neumann and the second R\'{e}nyi entropy of a bipartition of the infinite brick wall circuit into two halves are shown in Fig.~\ref{fig:Ising_transition2}. As before, the solid blue lines show the results of numerically simulating the circuit for different numbers of layers; the blue dots show the results from the quantum computer~\cite{IBMQ}. The peak in the entanglement entropies indicates the point of the phase transition, which lines up with the transition point seen in the magnetization. This shows that using a brick wall circuit as a variational ansatz still produces a single sharp transition point, where the nonanalytic features of the phase transition occur.

In fact, being able to observe all the relevant features of the phase transition using brick wall circuits was not obvious in advance. The circuit with its short correlation length being unable to reproduce the long-range correlations of the system around the critical point could have inhibited us from seeing the behavior at all. As it turns out, however, we can observe these features; the short correlation length of the circuit makes itself instead noticeable in the form of finite-depth scaling, which we will discuss now.

\section{Finite-Depth Scaling}
\label{sec:finite-depth_scaling}
In the previous section, we found that using brick wall circuits to approximate the ground state of the transverse-field Ising model along its phase transition shifts the transition point with increasing circuit depth (see Figs.~\ref{fig:Ising_transition1} and~\ref{fig:Ising_transition2}). This behavior is similar to the finite-entanglement scaling observed in MPSs~\cite{Tagliacozzo2008, Pollmann2009, Pirvu2012}, and correspondingly we refer to this behavior in brick wall circuits as \emph{finite-depth scaling}.

The finite-depth scaling in brick wall circuits stems from the finite correlation length of the ansatz. At the critical point, the correlation length of the true ground state diverges, which cannot be captured by the circuit. Instead, features of the transition are smoothed out, and the transition point is shifted. We can try to quantify the scaling behavior for the circuits analogously to how it was done for MPSs in Ref.~\cite{Tagliacozzo2008}. To do this, we assume the correlation length of the circuit at the critical point is related to the number of layers $m$ in the circuit by the scaling relation:
\begin{equation}
    \xi_m \sim m^{\lambda},
    \label{eq:xi_m}
\end{equation}
with some as of yet unknown constant $\lambda$. Note that $\lambda\leq1$ because, as discussed in the previous section, the correlation length can at most scale linearly with the circuit depth. For the time being, the assumption in Eq.\eqref{eq:xi_m} can be taken as an analogy to the case in MPSs, where the correlation length at the critical point scales with the bond dimension $\chi$ as $\xi_{\chi}\sim\chi^{\kappa}$~\cite{Tagliacozzo2008, Pollmann2009}.

Consider a Hamiltonian that depends on a parameter $t$ and has a phase transition at $t=0$. In terms of the Ising model discussed in the previous section, ${t = g-g_c}$, where $g$ is the strength of the transverse field, and $g_c=1$ is the location of the phase transition. Then the correlation length of the ground state diverges according to the power law:
\begin{equation}
    \xi \sim \left|t\right|^{-\nu},
    \label{eq:xi_t}
\end{equation}
as the critical point is approached, where $\nu$ is a critical exponent. By inverting this relation, we find
\begin{equation}
    \left|t\right| \sim \xi^{-1/\nu},
    \label{eq:t_xi}
\end{equation}
which effectively states how close we are to the critical point in terms of the correlation length. As the correlation length of the circuit cannot diverge, we cannot reach the actual critical point using the circuit. Instead, the critical point shifts to a pseudo critical point $t^*$, which is reached when the correlation length assumes its maximum. Inserting Eq.~\eqref{eq:xi_m} into Eq.~\eqref{eq:t_xi}, we find that the critical point shifts according to
\begin{equation}
    \left|t^*\right| \sim m^{-\lambda/\nu}.
    \label{eq:t_scaling}
\end{equation}

Consider a quantity $Q$ that diverges or vanishes at the critical point according to a universal exponent $\omega$, i.e., $Q \sim \left|t\right|^{\omega}$. This could, for example, be the order parameter of the system, which vanishes as the critical point is approached with the universal exponent $\omega = \beta$. For the Ising model, the order parameter is the magnetization, and $\beta=\frac{1}{8}$. Then Eqs.~\eqref{eq:xi_m} and~\eqref{eq:t_xi} tell us that this translates to a scaling behavior in terms of the circuit as
\begin{equation}
    Q \sim m^{-\lambda\omega/\nu}.
    \label{eq:Q_scaling}
\end{equation}

Apart from power law behaviors, there are also universal logarithmic divergences, such as for the entanglement entropy. The $n$th order R\'{e}nyi entropy of a bipartition of the system into two halves scales according to $S^{(n)}\sim\frac{c}{12}\left(1+\frac{1}{n}\right)\log\xi$, where $c$ is the central charge of the corresponding CFT~\cite{Holzhey1994, Calabrese2004, Calabrese2009}. In terms of the brick wall circuits, this translates to the scaling:
\begin{equation}
    S^{(n)}\sim\frac{c\lambda}{12}\left(1+\frac{1}{n}\right)\log\left(m\right).
    \label{eq:S_scaling}
\end{equation}

\section{Numerical Evidence of Finite-Depth Scaling}
\label{sec:numerical_evidence}
In this section, we numerically verify the finite-depth scaling discussed in the previous section for two different critical points. We consider the transverse-field Ising model with an additional symmetry-preserving term $-K Z_iZ_{i+1}$, which has central charge $c=\frac{1}{2}$, and the XXZ model with central charge $c=1$ as examples. Furthermore, we try to extract a numerical value for $\lambda$, which only appeared as an unknown constant until now.

\subsection{Quantum Ising model}
First, let us look at the transverse-field Ising model with the additional $-K Z_iZ_{i+1}$ term. The Hamiltonian then reads
\begin{equation}
    H = \sum_i -X_iX_{i+1} + g Z_i - K Z_iZ_{i+1}.
    \label{eq:TFI_K}
\end{equation}
As the added term respects the symmetry of the Ising model, varying the parameters $g$ and $K$ still leads to a phase transition between a symmetry-broken and a symmetric phase, whose universal exponents and central charge are that of the Ising transition. The critical points we will be considering lie at the following parameter pairs\footnote{For the standard transverse-field Ising model at $K=0$ the point of the phase transition is known. For $K\neq0$ we use the density matrix renormalization group (DMRG)~\cite{tenpy} to determine the point of the phase transition.}:
\begin{equation}
	\begin{tabular}{c|cccccc}
        \hline\hline
		$K$ & $0.0$ & $0.1$ & $0.2$ & $0.3$ & $0.4$ & $0.5$ \\ \hline
		$g$ & $1.000$ & $0.835$ & $0.680$ & $0.538$ & $0.409$ & $0.295$ \\ \hline\hline
	\end{tabular}.
	\label{eq:Kg_pairs}
\end{equation}

\begin{figure}
    \includegraphics[width=\linewidth]{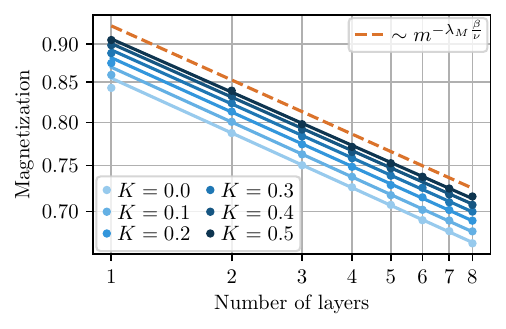}
    \caption{\textbf{Finite-depth scaling of the magnetization at the critical point of the Ising model with an additional $-K Z_iZ_{i+1}$ term.} The Hamiltonian of the system is $H = \sum_i\left(-X_iX_{i+1} + g Z_i - K Z_iZ_{i+1}\right)$. For each value of $K$, the strength of the transverse field $g$ is tuned such that the model is critical---see Eq.~\eqref{eq:Kg_pairs}. The blue dots show the magnetization of the circuit for different parameters; the solid blue lines show the results of fitting the expected power law in Eq.~\eqref{eq:Q_scaling} to the data. From the fitted values, we obtain an average value of $\lambda_M = 0.938 \pm 0.005$; for more details, see Appendix~\ref{app:data}. Inserting this value for $\lambda$ into Eq.~\eqref{eq:Q_scaling} yields the orange dashed line.}
    \label{fig:magnetization_scaling}
\end{figure}

\begin{figure*}
	\begin{minipage}[b]{0.49\linewidth}
		\includegraphics[height=0.67\linewidth]{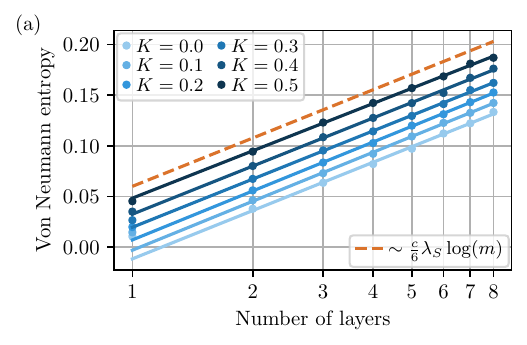}
	\end{minipage}
	\hfill
	\begin{minipage}[b]{0.49\linewidth}
        \includegraphics[height=0.67\linewidth]{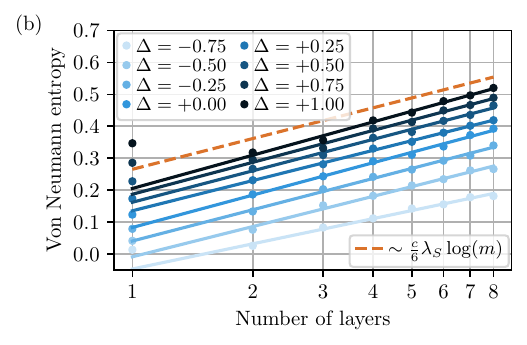}
	\end{minipage}
    \caption{\textbf{Finite-depth scaling of the entanglement entropy in (a) the Ising model with an additional $-K Z_iZ_{i+1}$ term and (b) the XXZ model.} The Hamiltonian of the Ising model with an additional $-K Z_iZ_{i+1}$ term is given in Eq.~\eqref{eq:TFI_K} and the parameters pairs in Eq.~\eqref{eq:Kg_pairs}; the Hamiltonian of the XXZ model is given in Eq.~\eqref{eq:XXZ}. The blue dots show the von Neumann entropy of a bipartition of the infinite circuit; the solid blue lines show the results of fitting the expected logarithmic behavior in Eq.~\eqref{eq:S_scaling} to the data. From the fitted values, we obtain an average (a) $\lambda_S = 0.826 \pm 0.004$ for the Ising model and (b) $\lambda_S = 0.834 \pm 0.024$ for the XXZ model; for more details, see Appendix~\ref{app:data}. The orange dashed lines show the logarithmic scaling with their respective values $\lambda_S$.}
    \label{fig:entanglement_scaling}
\end{figure*}

Let us take a look at the order parameter of the transition, which in this case is just the magnetization $\expval{X_i}$. According to Eq.~\eqref{eq:Q_scaling}, we expect the magnetization at the critical point to follow the power law $\expval{M}\sim m^{-\lambda\beta/\nu}$. This behavior is shown in Fig.~\ref{fig:magnetization_scaling}. The blue dots show the magnetization of the circuit for increasing circuit depth, and the different colors indicate the different critical points. The solid blue lines show the results of fitting a power law to the data, where the data points for a single layer have been excluded from the fit because the circuit is too shallow. We can see that, for every choice of $K$, the data points follow the power law nicely. More importantly, we can see that all the lines have the same slope, which is expected as all models have the same universal exponents. We can now extract $\lambda$ from the observed power law, as the values for the universal exponents are known for the Ising model---they are $\beta = \frac{1}{8}$ and $\nu = 1$. This gives an average value of $\lambda_M = 0.936 \pm 0.006$, which is close to the maximum possible value of $\lambda=1$. For more details on how we obtained the values of $\lambda$ and their uncertainties, see Appendix~\ref{app:data}.

We now turn to the finite-depth scaling of the entanglement entropy. The von Neumann entropy scales as $S\sim \frac{c\lambda}{6}\log m$ according to Eq.~\eqref{eq:S_scaling}, where the central charge $c=\frac{1}{2}$ for the Ising model. Figure~\ref{fig:entanglement_scaling}a shows the scaling behavior of the von Neumann entropy in the brick wall circuits for the Ising model. The blue dots show the average von Neumann entropy of a bipartition of the infinite circuit into two halves---once partitioned between two unit cells and once partitioned within a single unit cell---for the different critical points. The solid blue lines show the results of fitting the expected logarithmic scaling to the data, where again the data points for a single layer have been excluded as outliers from the fit because the circuit is too shallow. The data points all lie on a straight line in the figure, in accordance with the scaling relation in Eq.~\eqref{eq:S_scaling}. Moreover, Eq.~\eqref{eq:S_scaling} requires that all critical points exhibit a scaling with the same slope; this can also be seen to be fulfilled by the data in the figure. From the slope, we can calculate the value for $\lambda$, using $c=\frac{1}{2}$ for the Ising model, whose average comes out to be $\lambda_S = 0.824 \pm 0.006$. This result is close to but notably smaller than the value for $\lambda_M$ obtained from the magnetization before.

We can compare the entanglement scaling we found for infinite systems to the entanglement scaling of finite systems considered in Ref.~\cite{BravoPrieto2020}. In the finite-depth regime in finite systems, the size of the light cone is small compared with the system size. Consequently, we expect to find the same values for the logarithmic scaling in the finite-depth regime as for the infinite circuits. We can calculate $\lambda$ from the fitted data presented in Ref.~\cite{BravoPrieto2020} for the Ising model and obtain $\lambda = 0.78 \pm 0.24$. Within the uncertainty, this agrees with our result.

\subsection{XXZ model}
Let us now consider the XXZ model, whose Hamiltonian reads
\begin{equation}
    H = \sum_i X_iX_{i+1} + Y_iY_{i+1} + \Delta Z_iZ_{i+1}.
    \label{eq:XXZ}
\end{equation}
The parameter $\Delta$ controls the strength of the anisotropy in this model. For $-1<\Delta\leq1$, the XXZ model is critical and can be described as a Luttinger liquid~\cite{Giamarchi2004}. It has a central charge $c=1$~\cite{DiFrancesco1997}.

Figure~\ref{fig:entanglement_scaling}b shows the scaling of the von Neumann entropy in the XXZ model. As before for the Ising model, the blue dots show the half-chain von Neumann entropy of the circuit, and the solid blue lines show the results of fitting the expected logarithmic behavior to the data, excluding data points for a single layer as outliers. Note that, compared with the Ising model in Fig.~\ref{fig:entanglement_scaling}a, the entanglement entropy now has much larger values; this is because the central charge is now twice as large. The data points follow the expected logarithmic behavior closely, and all the slopes of the blue lines are about the same. This again conforms well with the behavior described by Eq.~\eqref{eq:S_scaling}. Using $c=1$, we can calculate $\lambda$ from the fitted values---this gives an average of $\lambda_S = 0.834 \pm 0.024$. The obtained result is very close to the value of $\lambda_S$ obtained for the Ising model; however, in general, we do not expect models with different central charges to have the same value for $\lambda$. We can compare our result for the scaling of entanglement entropy to the scaling found in Ref.~\cite{BravoPrieto2020} in the finite-depth regime for the finite-size XXZ model. Calculating $\lambda$ from the fitted data, we obtain $\lambda = 0.72 \pm 0.48$. This agrees with our findings within the uncertainty.

\section{Discussion}
\label{sec:discussion}
We have introduced brick wall circuits as a variational ansatz to represent ground states of infinite systems. By considering the transverse-field Ising model as an example, we have seen that, despite its simple structure, the circuit can capture relevant features of a phase transition. In this example, we have also seen that the point of the phase transition shifts with increased circuit depth---reminiscent of the finite-entanglement scaling in MPSs.

Based on these observations, we adapted the finite-entanglement scaling relations for MPSs to the finite-depth scaling of the brick wall circuits, introducing the parameter $\lambda$ that controls how the correlation length of the circuit scales with its depth, i.e., $\xi_m\sim m^{\lambda}$. We then examined the scaling behavior numerically on variations of the transverse-field Ising model and the XXZ model and found that the finite-depth scaling accurately describes the observed behavior. From these numerical examples, we could extract values for $\lambda$: from the scaling of the order parameter at the Ising transition, we obtained ${\lambda_M^{\text{Ising}} = 0.938 \pm 0.005}$, and from the scaling of the entropy, we obtained ${\lambda_S^{\text{Ising}} = 0.826 \pm 0.004}$ for the Ising model and ${\lambda_S^{\text{XXZ}} = 0.834 \pm 0.024}$ for the XXZ model. An open question remains: why is there a discrepancy between $\lambda_M^{\text{Ising}}$ and $\lambda_S^{\text{Ising}}$? Also, even though $\lambda_S^{\text{Ising}}$ and $\lambda_S^{\text{XXZ}}$ are very close, in general, we expect $\lambda$ to depend on the central charge of the model, as is the case for the finite-entanglement scaling of MPSs~\cite{Tagliacozzo2008, Pollmann2009}. How to describe this dependence more precisely is another open question.

Generally, the scaling relations we have studied here can be used to extrapolate information about the exact state from approximations with a finite-depth circuit. Once the value of $\lambda$ is known, these relations can also be used to extract critical exponents or the central charge of a system. For that, we would require an analytical formula for $\lambda$---just like in the case of finite-entanglement scaling for MPSs. There, a similar relation $\xi_{\chi}\sim\chi^{\kappa}$ exists, relating the bond dimension of the MPS to its correlation length at the critical point, and an analytical derivation for the value of $\kappa$ has been given in Refs.~\cite{Pollmann2009, Pirvu2012}.

An interesting task is the generalization of brick wall circuits to two dimensions, as that is where conventional tensor network methods struggle. The structure of the brick wall circuit can be straightforwardly adapted to two dimensions while keeping its light cone. It would therefore be interesting to see whether the scaling relations discussed here carry over to the two-dimensional case. In the near future, brick wall circuits might thus be a tool to study critical quantum many-body systems in two dimensions.

\begin{acknowledgments}
    We thank Sheng-Hsuan Lin for helpful discussions. The optimization of the brick wall circuits was implemented using the QGOpt library~\cite{qgopt}. Density matrix renormalization group calculations were performed using the TeNPy library~\cite{tenpy}. We acknowledge the Research Institute CODE of the Universität der Bundeswehr München for providing access to the IBM quantum computers. A.S. was supported by a Research Fellowship from the Royal Commission for the Exhibition of 1851. This paper was supported by the European Research Council under the European Union’s Horizon 2020 research and innovation program (Grant Agreement No. 771537). F.P. acknowledges the support of the Deutsche Forschungsgemeinschaft (German Research Foundation) under Germany’s Excellence Strategy EXC-2111-390814868. This paper is part of the Munich Quantum Valley, which is supported by the Bavarian state government with funds from the Hightech Agenda Bayern Plus.
\end{acknowledgments}

\bibliography{bibliography.bib}

\appendix
\section{Obtaining values for \texorpdfstring{$\lambda$}{lambda}}
\label{app:data}
In Sec.~\ref{sec:numerical_evidence}, we presented numerical evidence for the finite-depth scaling of a brick wall circuit and obtained numerical values for the parameter $\lambda$. Here, we will discuss in more detail how these values were obtained and how their uncertainty was calculated.

\begin{table}[b]
    \renewcommand{\arraystretch}{1.25}
    \setlength{\tabcolsep}{12pt}
    \centering
    \begin{tabular}{cc}
        \hline\hline
        $K$ & $\lambda$ \\ \hline
        $0.0$ & $0.949 \pm 0.006$ \\
        $0.1$ & $0.951 \pm 0.006$ \\
        $0.2$ & $0.942 \pm 0.007$ \\
        $0.3$ & $0.936 \pm 0.010$ \\
        $0.4$ & $0.928 \pm 0.018$ \\
        $0.5$ & $0.923 \pm 0.014$ \\ \hline\hline
    \end{tabular}
    \caption{\textbf{The values of $\lambda$ obtained from fitting the finite-depth scaling of the magnetization at the critical point of the Ising model with an additional $-K Z_iZ_{i+1}$ term for different values of $K$.} The Hamiltonian is given in Eq.~\eqref{eq:TFI_K_App}; $g$ is chosen such that the model is critical---see Eq.~\eqref{eq:Kg_pairs_App}. The uncertainty of $\lambda$ is calculated from the uncertainty of the parameters of the fit. The mean value of $\lambda$ is given by $\lambda = 0.938 \pm 0.005$, where the uncertainty is given by the standard error of the mean.}
    \label{tab:TFI_magnetization_fit}
\end{table}

First, we consider the scaling of the magnetization at critical points of the Ising model with an additional symmetry-preserving term, corresponding to Fig.~\ref{fig:magnetization_scaling}. As a reminder, the Hamiltonian of the model is
\begin{equation}
    H = \sum_i -X_iX_{i+1} + g Z_i - K Z_iZ_{i+1}
    \label{eq:TFI_K_App}
\end{equation}
and the parameter pairs of the critical points we consider are the following:
\begin{equation}
	\begin{tabular}{c|cccccc}
        \hline\hline
		$K$ & $0.0$ & $0.1$ & $0.2$ & $0.3$ & $0.4$ & $0.5$ \\ \hline
		$g$ & $1.000$ & $0.835$ & $0.680$ & $0.538$ & $0.409$ & $0.295$ \\ \hline\hline
	\end{tabular}.
	\label{eq:Kg_pairs_App}
\end{equation}
For each of those critical points we optimized circuits with up to eight layers to approximate the ground state. Calculating the magnetization $\expval{M} = \frac{1}{2}\expval{X_i + X_{i+1}}$, where we average over the two sites in the unit cell, yields the data points. According to Eq.~\eqref{eq:Q_scaling} we expect the magnetization to follow the scaling behavior
\begin{equation}
    \expval{M}\sim m^{-\lambda\beta/\nu},
\end{equation}
where the critical exponents $\nu=1$ and $\beta=\frac{1}{8}$ are known for the Ising phase transition. For every critical point, we can now fit a function of the form $a \cdot m^{-b}$ to the data and calculate ${\lambda = \frac{\nu}{\beta} b}$. The uncertainties of the parameters $a$ and $b$ of the fit can be obtained as the square root of the diagonal entries of their covariance matrix~\cite{scipy_curve_fit} and can be propagated to $\lambda$. Doing this, we obtain the values given in Table~\ref{tab:TFI_magnetization_fit}. The final value for $\lambda_M = 0.938\pm0.005$ we give in the main text is obtained by calculating the average of all values of $\lambda$ obtained for the different parameter pairs, and the uncertainty is given by the standard error of the mean. In contrast to the uncertainty in Table~\ref{tab:TFI_magnetization_fit}, which loosely speaking shows how good the fitted function describes the data points, the small standard error of the mean shows that the obtained values of $\lambda$ are all almost the same.

Next, we consider the scaling of the von Neumann entropy of the Ising model as presented in Fig.~\ref{fig:entanglement_scaling}a. The von Neumann entropy should follow the scaling in Eq.~\eqref{eq:S_scaling}, i.e.,
\begin{equation}
    S\sim\frac{c\lambda}{6}\log\left(m\right),
    \label{eq:S_scaling_App}
\end{equation}
with $c=\frac{1}{2}$ for the Ising transition. For every choice of parameters, we can now fit a function of the form $a\cdot\log(m)+b$ to the data and calculate $\lambda=\frac{6}{c}a$. The results of this are listed in Table~\ref{tab:TFI_entanglement_fit} where, as before, the uncertainty stems from the uncertainty in the fitted parameters. Calculating the average of all obtained values comes out to be $\lambda_S = 0.826 \pm 0.004$, where the uncertainty is given by the standard error of the mean.

\begin{table}[t]
    \renewcommand{\arraystretch}{1.25}
    \setlength{\tabcolsep}{12pt}
    \centering
    \begin{tabular}{cc}
        \hline\hline
        $K$ & $\lambda$ \\ \hline
        $0.0$ & $0.826 \pm 0.017$ \\
        $0.1$ & $0.837 \pm 0.009$ \\
        $0.2$ & $0.837 \pm 0.008$ \\
        $0.3$ & $0.824 \pm 0.011$ \\
        $0.4$ & $0.822 \pm 0.016$ \\
        $0.5$ & $0.809 \pm 0.011$ \\ \hline\hline
    \end{tabular}
    \caption{\textbf{The values of $\lambda$ obtained from fitting the finite-depth scaling of the von Neumann entropy at the critical point of the Ising model with an additional $-K Z_iZ_{i+1}$ term for different values of $K$.} The Hamiltonian is given in Eq.~\eqref{eq:TFI_K_App}; $g$ is chosen such that the model is critical---see Eq.~\eqref{eq:Kg_pairs_App}. The uncertainty of $\lambda$ is calculated from the uncertainty of the parameters of the fit. The mean value of $\lambda$ is given by $\lambda = 0.826 \pm 0.004$, where the uncertainty is given by the standard error of the mean.}
    \label{tab:TFI_entanglement_fit}
\end{table}

Finally, we consider the XXZ model with the Hamiltonian:
\begin{equation}
    H = \sum_i X_iX_{i+1} + Y_iY_{i+1} + \Delta Z_iZ_{i+1}.
    \label{eq:XXZ_App}
\end{equation}
The critical points we considered in the main text are ${\Delta\in\{-0.75,-0.50,-0.25,0.00,0.25,0.50,0.75,1.00\}}$. The scaling of the von Neumann entropy is shown in Fig.~\ref{fig:entanglement_scaling}b. As for the Ising model, the von Neumann entropy should follow the scaling in Eq.~\eqref{eq:S_scaling}:
\begin{equation}
    S\sim\frac{c\lambda}{6}\log(m),
\end{equation}
only that now the central charge takes a different value, $c=1$. As before, we can fit a logarithmic function ${a\cdot\log(m)+b}$ to the data and calculate $\lambda = \frac{6}{c}a$ for every choice of $\Delta$. The results are shown in Table~\ref{tab:XXZ_entanglement_fit}, where the uncertainty stems from the fitted parameters. The mean value and the standard error of the mean are $\lambda_S = 0.834 \pm 0.024$.

\begin{table}[t]
    \renewcommand{\arraystretch}{1.25}
    \setlength{\tabcolsep}{12pt}
    \centering
    \begin{tabular}{cc}
        \hline\hline
        $\Delta$ & $\lambda$ \\ \hline
        $-0.75$ & $0.68 \pm 0.03$ \\
        $-0.50$ & $0.82 \pm 0.04$ \\
        $-0.25$ & $0.85 \pm 0.03$ \\
        $\phantom{-}0.00$ & $0.88 \pm 0.03$ \\
        $\phantom{-}0.25$ & $0.82 \pm 0.02$ \\
        $\phantom{-}0.50$ & $0.85 \pm 0.04$ \\
        $\phantom{-}0.75$ & $0.87 \pm 0.04$ \\
        $\phantom{-}1.00$ & $0.90 \pm 0.04$ \\ \hline\hline
    \end{tabular}
    \caption{\textbf{The values of $\lambda$ obtained from fitting the finite-depth scaling of the von Neumann entropy at the critical point of the XXZ model for different values of $\Delta$.} The Hamiltonian is given in Eq.~\eqref{eq:XXZ}. The uncertainty of $\lambda$ is calculated from the uncertainty of the parameters of the fit. The mean value is given by $\lambda = 0.834 \pm 0.024$, where the uncertainty is given by the standard error of the mean.}
    \label{tab:XXZ_entanglement_fit}
\end{table}

\section{Obtaining data from the IBM quantum computers}
\label{app:error_correction}
In Sec.~\ref{sec:brick_wall_circuits}, we considered the phase transition of the transverse-field Ising model as an example to use brick wall circuits to approximate the ground state of a given model. We looked at the magnetization and the entanglement entropies (see Figs.~\ref{fig:Ising_transition1} and~\ref{fig:Ising_transition2}) to observe the transition. For these two observables, we also presented some data that were obtained on the IBM quantum computers~\cite{IBMQ}. First, we discuss how we measured the magnetization to obtain the data presented in Fig.~\ref{fig:Ising_transition1} and introduce the error mitigation techniques we use. Then we present several ways for measuring R\'{e}nyi entropies of brick wall circuits on a quantum computer and discuss how we obtained the data presented in Fig.~\ref{fig:Ising_transition2}.

\subsection{Measuring the magnetization}
\label{app:magnetization}
\begin{figure}[t]
    \centering
    \includegraphics[width=0.502\linewidth]{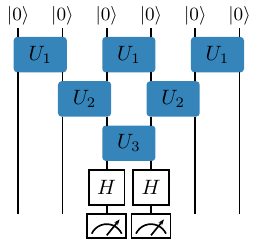}
    \caption{\textbf{The circuit used to measure the magnetization $\expval{M} = \frac{1}{2}\expval{X_i + X_{i+1}}$.} The circuit consists of all the gates of the brick wall circuit that fall within the light cone. Then two Hadamard gates are applied on the central qubits to rotate them into the $X$ basis, after which the measurement is performed on the two qubits.}
    \label{fig:magnetization_circuit}
\end{figure}

The magnetization $\expval{M} = \frac{1}{2}\expval{X_i + X_{i+1}}$ of a three-layer brick wall circuit can be measured with the circuit shown in Fig.~\ref{fig:magnetization_circuit}. The circuit consists of two parts. The first part comprises the blue gates, which make up the light cone of the circuit. The remaining gates of the infinite circuit do not contribute to the measurement and, as the circuit is invariant under translations by an even number, the magnetization does not depend on which two neighboring sites are chosen. To measure the magnetization of circuits with circuit depths other than $m=3$, the light cone simply needs to be adjusted accordingly. The second part of the circuit consists of the two Hadamard gates followed by measurements on the central two qubits. The Hadamards rotate the qubits into the $X$ basis, so that the subsequent measurement can be used to calculate the expectation value of the Pauli-$X$ operator. Denoting the number of shots on the quantum computer as $N$, the number of times we measure $\ket{00}$ as $n_{00}$ and the number of times we measure $\ket{11}$ as $n_{11}$, the magnetization can be obtained as $\expval{M} = \frac{n_{00}-n_{11}}{N}$. Note that the average magnetization for each of the results $\ket{01}$ and $\ket{10}$ is zero.

To obtain the results shown in Fig.~\ref{fig:Ising_transition1}, we ran the circuit with one and five layers on the IBM quantum computer named \texttt{montreal}. For each data point in the figure, we ran the circuit $42$ times in succession, with $\num{8192}$ shots per run. This gives us a total of $\num{344064}$ shots, which is more shots than the IBM quantum computer allows in a single run.

Following the measurement, we can correct the raw data from the quantum computer for two errors, namely, readout errors and global depolarizing errors. First, let us consider readout errors. These errors are bit-flip errors that occur when the qubit is measured and assigned the wrong output to the classical bit. A simple way to mitigate this error is readily implemented in Qiskit~\cite{qiskit}. This consists of constructing a readout matrix that gives the probability of measuring a given basis state as another basis state by running a calibration circuit on the quantum computer, which prepares every basis state. Then we can apply an appropriate pseudo-inverse of this readout matrix to the measurement outcome of the real experiment to correct for the average readout error.

We also correct the data from the quantum computer for global depolarizing errors, as outlined in Ref.~\cite{Vovrosh2021}. The basic idea is that the state prepared on the quantum computer is not described by the density matrix $\rho_0$ of the applied circuit but instead---due to global depolarizing errors---by the density matrix $\Tilde{\rho} = \left(1-p\right)\rho_0 + \frac{p}{2^N}\Id$; here, $p$ denotes the strength of the deviation from the expected density matrix, $\Id$ is the identity matrix, and $N$ is the number of qubits. When measuring an observable $O$ on the quantum computer, we obtain the expectation value with respect to the perturbed density matrix $\Tilde{\rho}$, i.e.,
\begin{equation}
    \Tr\left(O\Tilde{\rho}\right) = \left(1-p\right)\Tr\left(O\rho_0\right) + \frac{p}{2^N}\Tr O.
    \label{eq:perturbed_exp}
\end{equation}
Thus, if we were to know $p$, then we could deduce the expectation value with respect to the unperturbed density matrix as
\begin{equation}
    \Tr\left(O\rho_0\right) = \frac{\Tr\left(O\Tilde{\rho}\right)-\frac{p}{2^N}\Tr O}{\left(1-p\right)}.
    \label{eq:unperturbed_exp}
\end{equation}
In practice, $p$ can be calculated from a measurement where the expected outcome is known. Then for measurements with a similar circuit setup, $p$ can be assumed to be the same. In our case the observable $O$ is the magnetization ${M = \frac{1}{2}\left(X_i + X_{i+1}\right)}$, which is traceless, and hence, Eq.~\eqref{eq:perturbed_exp} simply becomes
\begin{equation}
    \Tr\left(M\Tilde{\rho}\right) = \left(1-p\right)\Tr\left(M\rho_0\right).
\end{equation}
Using this relation, we can calculate $p$ for one value of $g$ where the expected magnetization is known and then correct the data for all other values of $g$. An obvious choice to calculate $p$ would be for $g=0$ because, there, it is known that the circuit has magnetization $\expval{M} = 1$. However, at that point, the circuit consists only of single-qubit gates and thus has a different structure from the circuit for other values of $g$, leading to a different value for $p$. Since we already know the results for the magnetization from simulating the circuits on a classical computer, we could in principle choose any other value of $g$ to calculate $p$. Here, we choose $g=0.1$ which corresponds to the second data point in Fig.~\ref{fig:Ising_transition1}. Note that the perturbation $\frac{p}{2^N}\Id$ of the expected density matrix cannot produce a nonzero magnetization, and thus cannot be the reason the magnetization deviates from zero if the expected value would be zero. Therefore, we can only apply this mitigation scheme in the regime where the magnetization is nonzero.

\subsection{Measuring R\'{e}nyi entropies of brick wall circuits}
\label{app:renyi_entropies}
In this section, we present three different ways to measure the R\'{e}nyi entropies of a bipartition of the infinite brick wall circuit into two halves. We also give more details about how we obtained the data presented in Fig.~\ref{fig:Ising_transition2}. As a reminder, the R\'{e}nyi entropy of order $n$ is defined as $S^{(n)} = \frac{1}{1-n}\log\Tr\rho^n$, where $\rho$ is the reduced density matrix of the subsystem, so in our case $\rho$ is the reduced density matrix of one half of the infinite state. All methods we present here will give some way to calculate $\Tr\rho^n$ on the quantum computer.

\begin{figure}[t]
    \centering
    \includegraphics[width=0.971\linewidth]{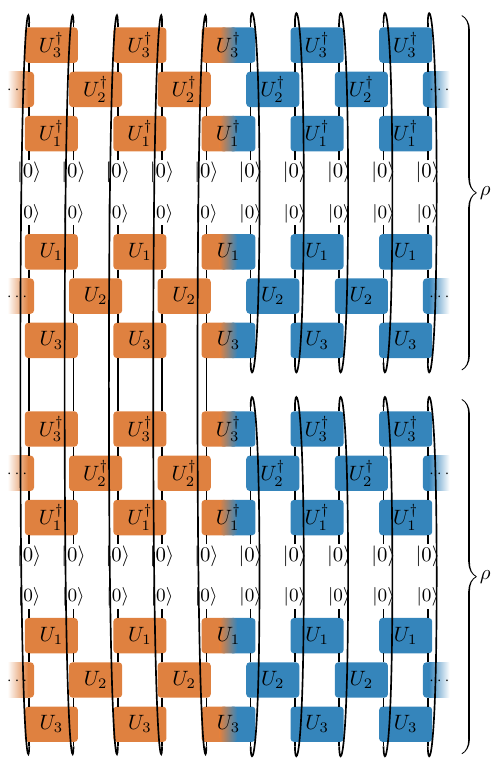}
    \caption{\textbf{Calculating $\Tr\rho^2$ for one half of an infinite brick wall circuit.} The brick wall circuit is partitioned into two halves, the left half shown in orange and the right half shown in blue. Contracting the outgoing quantum wires in the blue half with those of the conjugate circuit gives the reduced density matrix $\rho$ of the orange half, indicated in the figure by the braces. Contracting the two copies of the density matrix gives $\Tr\rho^2$. To calculate $\Tr\rho^n$ for any integer $n$, one can simply add more copies of the density matrix.}
    \label{fig:reduced_density_matrix}
\end{figure}

\begin{figure}[t]
    \centering
    \includegraphics[width=\linewidth]{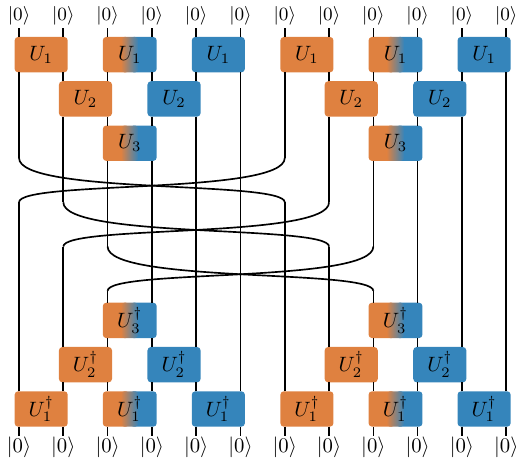}
    \caption{\textbf{Simplified circuit for calculating $\Tr\rho^2$ for one half of an infinite brick wall circuit.} Simplifying the circuit in Fig.~\ref{fig:reduced_density_matrix} by canceling unitary gates with their adjoints and arranging the gate blocks differently, we arrive at the above circuit. This circuit can be run on a quantum computer.}
    \label{fig:renyi_circuit}
\end{figure}

The first approach is to directly calculate $\Tr\rho^n$ on the quantum computer by contracting the density matrices. Consider a brick wall circuit that is partitioned into two parts, as shown in Fig.~\ref{fig:reduced_density_matrix}. The left half of the circuit is colored orange, and the right half of the circuit is colored blue. By contracting the outgoing quantum wires of the circuit in the blue half with those of the conjugate circuit, we obtain the reduced density matrix $\rho$ of the orange half, which is indicated by the curly brace. Taking two copies of the reduced density matrix and contracting them as shown in the figure yields $\Tr\rho^2$. We can calculate $\Tr\rho^n$ for any integer $n$ this way, by contracting $n$ copies of the reduced density matrix. The expression shown in Fig.~\ref{fig:reduced_density_matrix} can be further simplified by canceling every unitary gate that is contracted with its adjoint, leaving behind only the light cones along every cut of the bipartition into two halves. Reordering the remaining circuit blocks, we arrive at the circuit in Fig.~\ref{fig:renyi_circuit}. This circuit can be run directly on the quantum computer. Note that Ref.~\cite{Gopalakrishnan2019} already pointed out, in the context of numerical MPS calculations, that the reduced density matrix of the circuit can be written this way in terms of its gates. This method can be used to directly compute $\Tr\rho^n$ on a quantum computer. To obtain the R\'{e}nyi entropy, we simply run the circuit $N$ times and count the number of times $n_0$ that all qubits are in the $\ket{0}$ state after the measurement. Then the R\'{e}nyi entropy is given by
\begin{equation}
    S^{(n)} = \frac{1}{1-n} \log\sqrt{\frac{n_0}{N}}.
\end{equation}

\begin{figure}[t]
    \centering
    \includegraphics[width=\linewidth]{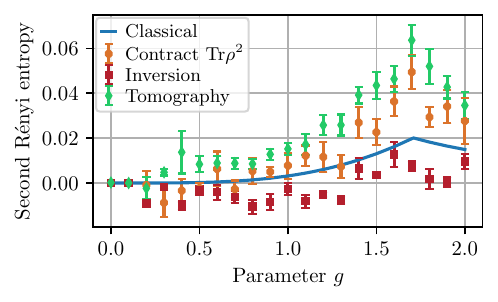}
    \caption{\textbf{Comparing different ways to measure the second R\'{e}nyi entropy.} We consider the second R\'{e}nyi entropy of a single-layer brick wall circuit approximating the ground state of the transverse-field Ising model ${H = \sum_i \left(-X_iX_{i+1} + g Z_i\right)}$---see also Fig.~\ref{fig:Ising_transition2}b. The solid blue line shows the result from simulating the circuit on a classical computer. The orange data points were obtained by using a circuit that contracts the reduced density matrices, the red data points were obtained via inversion symmetry, and the green data points were obtained via state tomography. For more details on the different methods, see the main text. The error bars are the standard error of the mean of several runs.}
    \label{fig:comparing_renyi}
\end{figure}

The data shown for the second R\'{e}nyi entropy in Fig.~\ref{fig:Ising_transition2} in the main text were obtained using the method described above. For each data point in the figure, we ran the circuit on the IBM quantum computers named \texttt{montreal} and \texttt{hanoi}. On \texttt{montreal}, we ran each circuit $42$ times in succession with $\num{32000}$ shots each run, giving the equivalent of $\num{1344000}$ shots. On \texttt{hanoi}, we ran each circuit $14$ times in succession with $\num{100000}$ runs shots per run, giving the equivalent of $\num{1400000}$ shots. We did this procedure four times on \texttt{montreal} and twice on \texttt{hanoi}; the average of the six results is presented in Fig.~\ref{fig:Ising_transition2}; the error bars show the standard error of the mean. The data are also presented in Fig.~\ref{fig:comparing_renyi} in orange. Additionally, data for the other two methods we will discuss in this section are shown in red and green. The exact results from simulating the circuit on a classical computer are shown in blue.

The raw data from the quantum computer are corrected for readout errors~\cite{qiskit} and global depolarizing errors~\cite{Vovrosh2021}. For the readout error correction, the same technique is applied as was previously used for the measurement of the magnetization. The correction of global depolarizing errors follows the same ideas as before but needs to be slightly adapted. Since we are counting the number of times that the final state after the measurement is $\ket{0\dots0}$, we are essentially measuring the operator $O = \ket{0\dots0}\bra{0\dots0}$. With this, Eq.~\eqref{eq:perturbed_exp} becomes
\begin{equation}
    \bra{0\dots0}\Tilde{\rho}\ket{0\dots0} = \left(1-p\right) \bra{0\dots0}\rho_0\ket{0\dots0} + \frac{p}{2^N}.
\end{equation}
Note that, here, $\rho_0$ refers to the density matrix of the state prepared on the quantum computer before the final measurement and not to the reduced density matrix $\rho$ of a bipartition of the brick wall circuit. From the above relation, we can calculate $p$ for $g=0.1$ and then correct the results for every other value $g>0.1$. For $g=0$, the state only consists of single qubit gates, and no mitigation of global depolarizing errors is needed.

\begin{figure}[t]
    \centering
    \includegraphics[width=\linewidth]{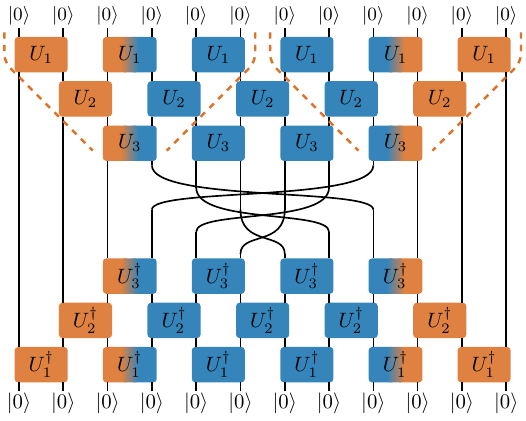}
    \caption{\textbf{Calculating $\Tr\rho^2$ via inversion symmetry as discussed in Ref.~\cite{Pollmann2012}.} The inversion symmetry is applied to the blue part of the circuit, which is long enough so that the light cones of the edges, shown as orange dashed lines, do not overlap. The two edges are therefore decoupled, and the resulting R\'{e}nyi entropy is precisely twice that of the infinite half-chain entropy.}
    \label{fig:inversion_circuit}
\end{figure}

Another way to calculate the second R\'{e}nyi entropy is given in Ref.~\cite{Pollmann2012}. If a state is inversion symmetric, then applying the inversion symmetry to a large region of the state yields $\pm\Tr\rho^2$, where $\rho$ is the reduced density matrix of that region, and the sign is related to the topological phase of the state. For a brick wall circuit, this situation is shown in Fig.~\ref{fig:inversion_circuit}. There, the inversion symmetry is applied to the central region colored in blue. Note that, instead of a bipartition into two halves with a single boundary between the two regions, we now have a bipartition into two regions with two boundaries. To obtain the entropy of a single boundary, we must make the region the inversion symmetry is applied to large enough such that the light cones of its boundaries do not overlap---as is shown in the figure. Then the boundaries decouple, and the entropy obtained from the circuit is just twice that of a single boundary. Thus, on the quantum computer, we can run the circuit $N$ times and count the number of times $n_0$ that the state after the measurement is $\ket{0\dots0}$, to obtain the second R\'{e}nyi entropy:
\begin{equation}
    S^{(2)} = - \frac{1}{2} \log\sqrt{\frac{n_0}{N}}.
\end{equation}
The factor $\frac{1}{2}$ appears to account for the two boundaries.

The data from this method are presented as red squares in Fig.~\ref{fig:comparing_renyi}. To obtain the data, we ran the circuits on the IBM quantum computers named \texttt{montreal} and \texttt{hanoi}. On \texttt{montreal}, each circuit ran $42$ times in succession with $\num{32000}$ shots each run, giving the equivalent data of a run with $\num{1344000}$ shots. On \texttt{hanoi}, each circuit ran $14$ times in succession with $\num{100000}$ shots per run, giving data equivalent to a run with $\num{1400000}$ shots. We did this procedure four times on \texttt{montreal} and twice on \texttt{hanoi}; the average of the six results is presented in Fig.~\ref{fig:comparing_renyi}; the error bars show the standard error of the mean.

Again, we corrected the raw data from the quantum computer for readout errors~\cite{qiskit} and global depolarizing errors~\cite{Vovrosh2021} with the same methods as before. Equation~\eqref{eq:perturbed_exp} for the correction of global depolarizing errors becomes
\begin{equation}
    \bra{0\dots0}\Tilde{\rho}\ket{0\dots0} = \left(1-p\right) \bra{0\dots0}\rho_0\ket{0\dots0} + \frac{p}{2^N},
\end{equation}
from which we can calculate $p$ for $g=0.1$ and then can correct the data for all data points with $g>0.1$.

\begin{figure}[t]
    \centering
    \includegraphics[width=0.502\linewidth]{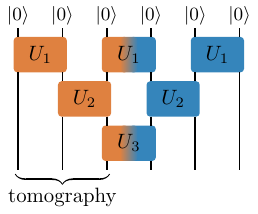}
    \caption{\textbf{The light cone of the brick wall circuit needed for state tomography.} Performing state tomography on the left half of the circuit colored in orange yields a density matrix that can be used to calculate the half-chain entanglement entropy of the infinite state.}
    \label{fig:tomography}
\end{figure}

Finally, we can also obtain the reduced density matrix by state tomography. This is a procedure that is readily implemented in Qiskit~\cite{qiskit}. State tomography performs a set of measurements in different bases on a subset of qubits of the state and then can reconstruct the reduced density matrix of the subsystem. The number of different measurements needed scales exponentially with the size of the subsystem on which state tomography is performed, so it becomes too costly for very large subsystems. However, for small systems, this works very well. To calculate the half-chain R\'{e}nyi entropy of a brick wall circuit with state tomography, it is enough to consider the light cone of the circuit along the cut into two subsystems, as shown in Fig.~\ref{fig:tomography}. This is because, for calculating the R\'{e}nyi entropy of the state, we do not actually need the density matrix $\rho$ of the half-infinite state; the reduced density matrix $\rho'$ of the orange half of the light cone in Fig.~\ref{fig:tomography} suffices. When calculating $\Tr\rho^n$ (see Fig.~\ref{fig:reduced_density_matrix} for an example with $n=2$), all gates in the blue half that are not part of the light cone cancel with the adjoint of the state after tracing out the blue half of the state, and all gates in the orange half that are not part of the light cone cancel with the adjoint gates of the copy of $\rho$. After the cancellation, only the light cones remain, and we effectively calculate $\Tr\rho'^n$ (see Fig.~\ref{fig:renyi_circuit} for an example with $n=2$). Thus, performing state tomography on the orange subsystem in Fig.~\ref{fig:tomography} yields a density matrix $\rho'$ that we can use to calculate the $n$th order R\'{e}nyi entropy:
\begin{equation}
    S^{(n)} = \frac{1}{1-n} \log\Tr\rho'^n,
\end{equation}
which gives the same result for the entropy as the reduced density matrix $\rho$ of the half-infinite state. Note, however, that in general $\rho'\neq\rho$, and so calculating other observables with the reduced density matrix obtained from state tomography will yield different results.

We show the results of state tomography in Fig.~\ref{fig:comparing_renyi} along with data from the previously discussed methods. To obtain these results, we ran the tomography circuits on the IBM quantum computer \texttt{montreal} six times, with $\num{32000}$ shots for each measurement. The data shown are the mean of the six runs, with the error bars showing the standard error of the mean.

The data from the quantum computer are, as before, corrected for readout errors~\cite{qiskit} and global depolarizing errors~\cite{Vovrosh2021}, as discussed before. The method for correcting global depolarizing errors needs to be adapted slightly to the case at hand. Remember that, instead of the density matrix of the circuit $\rho_0$, on the quantum computer, we actually construct the perturbed density matrix ${\Tilde{\rho} = \left(1-p\right)\rho_0 + \frac{p}{2^N}\Id}$---hence, $\Tilde{\rho}$ is the density matrix that will be reconstructed by state tomography and not $\rho_0$. Calculating $\Tr\Tilde{\rho}^2$, which we need to get the second R\'{e}nyi entropy, we obtain the relation:
\begin{equation}
    \Tr\Tilde{\rho}^2 = \left(1-p\right)^2 \Tr\rho_0^2 + \frac{p}{2^{N-1}} - \frac{p^2}{2^N}.
\end{equation}
From this relation, we can again calculate $p$ for $g=0.1$ and subsequently obtain the unperturbed $\Tr\rho_0^2$ from the data obtained on the quantum computer for all $g>0.1$ by assuming a constant $p$.

Another way to measure R\'{e}nyi entropies that we have not yet considered so far is to use randomized measurements~\cite{vanEnk2012, Elben2018, Vermersch2018, Brydges2019, Elben2019}. From the statistical correlations of measurements after applying random unitary gates to a subsystem, the R\'{e}nyi entropies can be inferred. Like state tomography, it would be enough to consider the light cone of the circuit in Fig.~\ref{fig:tomography} to obtain the infinite-half chain entanglement entropy, but instead of performing state tomography on the orange half of the system, we could apply random unitaries before the measurement.

Note that all methods we have presented for measuring R\'{e}nyi entropies, as well as the data presented in Fig.~\ref{fig:comparing_renyi}, implied that we cut the circuit into two parts within the unit cell. All presented methods work analogously for the case where we cut the circuit into two parts between two unit cells. This leads effectively to considering the circuit with its final layer removed and cutting that circuit within its unit cell.
\end{document}